\documentclass[preprint,aps,a4paper,amsfonts,amsmath,amssymb,superscriptaddress,groupedaddress,nofootinbib]{revtex4}
\pdfoutput=1
\usepackage{makeidx}
\usepackage{graphicx}

\usepackage{ulem}
\usepackage{color}

\begin{document}

\title{Statistical analysis and stochastic interest rate modelling for valuing the future with implications in climate change mitigation}

\author{Josep Perell\'o}
\email{josep.perello@ub.edu}
\affiliation{Department of Condensed Matter Physics, University of Barcelona, Catalonia, Spain}
\affiliation{Institute of Complex Systems (UBICS), University of Barcelona, Catalonia, Spain}

\author{Miquel Montero}
\affiliation{Department of Condensed Matter Physics, University of Barcelona, Catalonia, Spain}
\affiliation{Institute of Complex Systems (UBICS), University of Barcelona, Catalonia, Spain}

\author{Jaume Masoliver} 
\affiliation{Department of Condensed Matter Physics, University of Barcelona, Catalonia, Spain}
\affiliation{Institute of Complex Systems (UBICS), University of Barcelona, Catalonia, Spain}

\author{J. Doyne Farmer}
\affiliation{Mathematical Institute and Institute for New Economic Thinking at the Oxford Martin School, University of Oxford, Oxford, UK}
\affiliation{Santa Fe Institute, Santa Fe, New Mexico, USA}

\author{John Geanakoplos}
\affiliation{Santa Fe Institute, Santa Fe, New Mexico, USA}
\affiliation{Department of Economics, Yale University, New Haven, CT, USA}

\date{\today}

\begin{abstract}
High future discounting rates favor inaction on present expending while lower rates advise for a more immediate political action. A possible approach to this key issue in global economy is to take historical time series for nominal interest rates and inflation, and to construct then real interest rates and finally obtaining the resulting discount rate according to a specific stochastic model. Extended periods of negative real interest rates, in which inflation dominates over nominal rates, are commonly observed, occurring in many epochs and in all countries. This feature leads us to choose a well-known model in statistical physics, the Ornstein-Uhlenbeck model, as a basic dynamical tool in which real interest rates randomly fluctuate and can become negative, even if they tend to revert to a positive mean value. By covering 14 countries over hundreds of years we suggest different scenarios and include an error analysis in order to consider the impact of statistical uncertainty in our results. We find that only 4 of the countries have positive long-run discount rates while the other ten countries have negative rates. Even if one rejects the countries where hyperinflation has occurred, our results support the need to consider low discounting rates. The results provided by these fourteen countries significantly increase the priority of confronting global actions such as climate change mitigation. We finally extend the analysis by first allowing for fluctuations of the mean level in the Ornstein-Uhlenbeck model and secondly by considering modified versions of the Feller and lognormal models. In both cases, results remain basically unchanged thus demonstrating the robustness of the results presented.
\end{abstract}

\maketitle

\section{Introduction}

Statistical physics have been paying attention to economics and finance by providing new models and analyzing data available \cite{Mantegna1999,Bardoscia2017,Bouchaud2019}. Most of the contributions investigate the nature of financial markets based on historical records, even its microstructure (see e.g. \cite{Bouchaud2019b,Farmer2005}) or alternatively from a rather macroscopic and aggregated level (see e.g. \cite{Masoliver2002,Perello2003,Perello2004a,Perello2004,Masoliver2006,Perello2006,Perello2008,Camprodon2012,Bormetti2008,Delpini2011,Delpini2015}). However, there are still several issues in which an approach from physics can offer new perspectives and results. This is, for instance, the case of ``discounting" which in economics refers to weighting the future relative to the present \cite{Samuelson}. Discounting constitutes the subject of this paper.

The choice of a discounting function has enormous consequences in many aspects of the global economy as, for instance, long-run environmental planning and, more specifically, climate action \cite{DasGupta2004}. In a highly influential report on climate change commissioned by the UK government, Stern \cite{Stern} uses a discounting rate of $1.4\%$ while Nordhaus \cite{Nordhaus} argues for a discount rate of $4\%$ and at other times \cite{Nordhaus2007} has advocated rates as high as $6\%$. Both estimates constitute a completely different point of view on how to address climate change. Indeed, while Stern's estimate would imply immediate spending, Nordhaus's figures indicate that immediate and strong action would be unnecessary. The choice of discount rate is, therefore, one of the biggest factors influencing the debate on the urgency of the response to climate change. Although Stern has been widely criticized for using such a low rate \cite{Nordhaus,Nordhaus2007,DasGupta2006,Mendelsohn,Weitzman2007,Nordhaus2008}, our estimates are on average much closer to Stern than to Nordhauss and support more substantial immediate spending on climate actions. The Calderon report in July 2014 has also claimed that there is a false dilemma behind the choice between the economy growth and the environmental responsibility \cite{Stern2015,new_climate}.
  
Economists present a variety of reasons for discounting, including impatience, economic growth, and declining marginal utility; these are embedded in the Ramsey formula, which forms the basis for the standard approaches to discounting \cite{ArrowReview,Chichilnisky2018}. Here we adopt the net present value approach, which treats the real interest rate as the measure of the trade-off between consumption today and consumption next year, without delving into the factors influencing the real interest rate.

It is often argued that, based on past trends in economic growth, future technologies will be so powerful compared with present technologies that it is more cost-effective to
encourage economic growth --or solving other problems such as AIDS or malaria-- than it is to take action against global warming now \cite{Nordhaus2008}. Analyses supporting this conclusion typically study discounting by working with an interest rate that is fixed over time, ignoring fluctuations about the average. This is mathematically convenient, but it is also dangerous: In this problem, as in many others, fluctuations play a decisive role.

A proper analysis takes fluctuations in the real interest rate, caused partly by fluctuations in growth, into account \cite{Weitzman98,Gollier,NewellPizer}. When the real interest rate $r(t)$ varies randomly the discounting function becomes \cite{farmer2015}
\begin{equation}
D(t) = \mathbb E \left[ \exp\left( {-\int_{0}^{t} r(t^{\prime})dt^{\prime}}\right)\right],
\label{D}
\end{equation}
where the expectation $\mathbb E[\cdot]$ is an average over all possible interest rate paths. The fact that this is an average of exponentials, and not an exponential of an average, implies that the paths with the lowest interest rates dominate. This has been shown in several ways. Early papers analyzed an extreme case in which the annual real rate is unknown today, but starting tomorrow it will be fixed forever at one of a finite number of values \cite{Weitzman98,Gollier}. Other papers simulate stochastic interest rate processes out to some horizon, leaving aside the asymptotic behavior of real rates \cite{NewellPizer,Groom,Hepburn,Freeman}. 

The presence of fluctuations can dramatically alter the functional form of the discounting function. If real interest rates follow a geometric random walk, for example, the discounting function asymptotically may decay as a power law of the form $D(t) = At^{-1/2}$ \cite{Farmer} (see Sect. VI). In contrast to the exponential function, this is not integrable on $(0, \infty)$, underscoring how important the effect of persistent fluctuations can be. We have recently analyzed these issues by considering three of the most popular stochastic models for the dynamics of interest rates \cite{farmer2015}: Ornstein-Uhlenbeck \cite{Uhlenbeck1930}, Feller \cite{Feller1951}, and lognormal \cite{Osborne1959} processes, which are also very relevant in statistical physics. The Ornstein-Uhlenbeck (OU) model \cite{Uhlenbeck1930} is the only one that allows for negative rates $r<0$ and its asymptotic expression has an exponential decay with a long-run rate $r_\infty$ that  differs from historical average interest rates by being substantially smaller, zero or eventually negative. We here want to go one step further and provide empirical estimates to such a discount based on historical data of interest rates from Argentina, Australia, Canada, Chile, Denmark, Germany, Italy, Japan, Netherlands, South Africa, Spain, Sweden, United Kingdom, and the United States. Such a diversity of countries, representing a variety of scenarios, allows us to better explore the intrinsic randomness of the real interest rates and how they lead to different costs of global economy planning such as climate action.


\section{Building real interest rates with the empirical data available}

Real interest rates are nominal rates corrected by inflation so we need first of all to study nominal rates and inflation separately. The countries in our sample are: Argentina (ARG, 1864-1960), Australia (AUS, 1861-2012), Canada (CAN, 193-2012), Chile (CHL, 1925-2012), Denmark (DNK, 1821- 2012), Germany (DEU, 1820-2012), Italy (ITA, 1861-2012), Japan (JPN, 1921-2012), Netherlands (NLD, 1813-2012), South Africa (ZAF, 1920-2012), Spain (ESP, 1821-2012), Sweden (SWE, 1868-2012), United Kingdom (GBR, 1694-2012), and the United States (USA, 1820-2012). The details of each sample are reported in Table \ref{table1}.

\begingroup
\squeezetable
\begin{table}[!ht]
\centering
\caption{
{\bf Description of the empirical data.} Each column represents the data source from 14 different countries with their time periods and  frequencies. The number of records corresponds to the resulting real interest rate historical time series.\label{table1}}
\begin{tabular}{clllccc}
\hline\hline
	&	Country	& Consumer Price Index & Bond Yields &	from	&	to 	&	\# records\\\hline
1 	&	Argentina	&CPARGM	&IGARGM		&12/31/1864	&03/31/1960	&342\\
	&			&annual from 12/31/1864&quarterly\\
	&			& quarterly from 12/31/1932\\
2	&	Australia		&CPAUSM	&IGAUS10		&12/31/1861	&09/30/2012	&564\\
	&			&annual from 12/31/1861	&quarterly\\
	&			& quarterly 12/31/1991\\
3	&	Canada	&CPCANM	&IGCAN10	&12/31/1913	&09/30/2012	&357	\\	
	&			&quarterly	&quarterly	\\
4	&	Chile		&CPCHLM	&IDCHLM$^{1}$ &03/31/1925	&09/30/2012	&312\\	
	&			&quarterly	&quarterly			\\
5	&	Denmark		&CPDNKM	&IGDNK10		&12/31/1821	&09/30/2012	&725\\
	&			&annual from 12/31/1821	&quarterly\\
	&			&quarterly from 12/31/1914\\	
6	&	Germany	&CPDEUM	&IGDEU10$^{2}$	&12/31/1820	&09/30/2012	&729\\
         &			&annual from 12/31/1820	&quarterly	\\
	&			&quarterly from 12/31/1869	\\
7	&	Italy		&CPITAM	&IGITA10	&12/31/1861	&09/30/2012	&565	\\
	&			&annual from 12/31/1861&quarterly\\
	&			&quarterly from 12/31/1919\\ 	
8	&Japan		&CPJPNM	&IGJPN10D$^{6}$	&12/31/1921	&12/31/2012	&325\\
	&			&quarterly	&quarterly\\
9	& Netherlands	&CPNLDM	&IGNLD10D$^{5}$	&12/31/1813	&12/31/2012	&189\\
	&			&annual	&annual\\
10	& South Africa	&CPZAFM	&IGZAF10		&12/31/1920	&09/30/2012	&329\\
				&&quarterly	&quarterly	\\
11	& Spain$^{3}$	&CPESPM	&IGESP10$^{4}$	&12/31/1821	&09/30/2012	&709\\
	&			& annual from 12/31/1821	&quarterly\\
	&			& quarterly from 12/31/1920\\
12	& Sweden		&CPSWEM	&IGSWE10		&12/31/1868	&09/30/2012	&135\\	
	&			&annual	&annual\\
13	&United Kingdom&CPGBRM	&IDGBRD$^{1}$	&12/31/1694	&12/31/2012	&309\\	
	&			&annual	&annual\\
14	&United States	&CPUSAM	&TRUSG10M	&12/31/1820	&10/30/2012	&183\\
	&			&annual	&annual\\
\hline\hline
\end{tabular}
\footnotetext{($^1$) We have taken the Discount (ID) rate since the Government Bond Yield data was not available. ($^2$) From 06/30/1915 to 03/31/1916 IGDEU is empty and we have repeated the previous record. ($^3$) From 07/31/1936 to 12/31/1940 no records available. ($^4$) 07/31/1936 is empty and we have repeated the previous record. ($^5$) 12/31/1945 is empty and we have repeated the previous record. ($^6$) From 12/31/1946 to 09/30/1948 is empty and we have repeated the previous record.}
\end{table}
\endgroup


Nominal rates can be obtained through the 10 year Government Bond Yield (see Table \ref{table1} for further details). Following the standard procedure provided by the literature (see, for instance, \cite{Brigo2006}), we transform the annual rate $\beta(t|T)$, where $T=10$ years, into logarithmic rates, and denote the resulting nominal rates time series by 
$$
n(t)=\ln[1+\beta(t|T)].
$$
 
The inflation rate $i(t)$ is estimated through the Consumer Price Index (CPI) $C(t)$ by
$$
i(t)=\frac{1}{T}\sum_{j=0}^{T-1} \ln\left[1+C(t+j)\right]
$$
where $T$ is chosen to be 10 years to be consistent with the 10 year nominal rate. We have, therefore, smoothed inflation rates with a ten-year forward moving average as this is again the standard procedure in these cases. 

\begin{figure}[!h]
\vspace*{.05in}
\centering
\includegraphics[width=13cm]{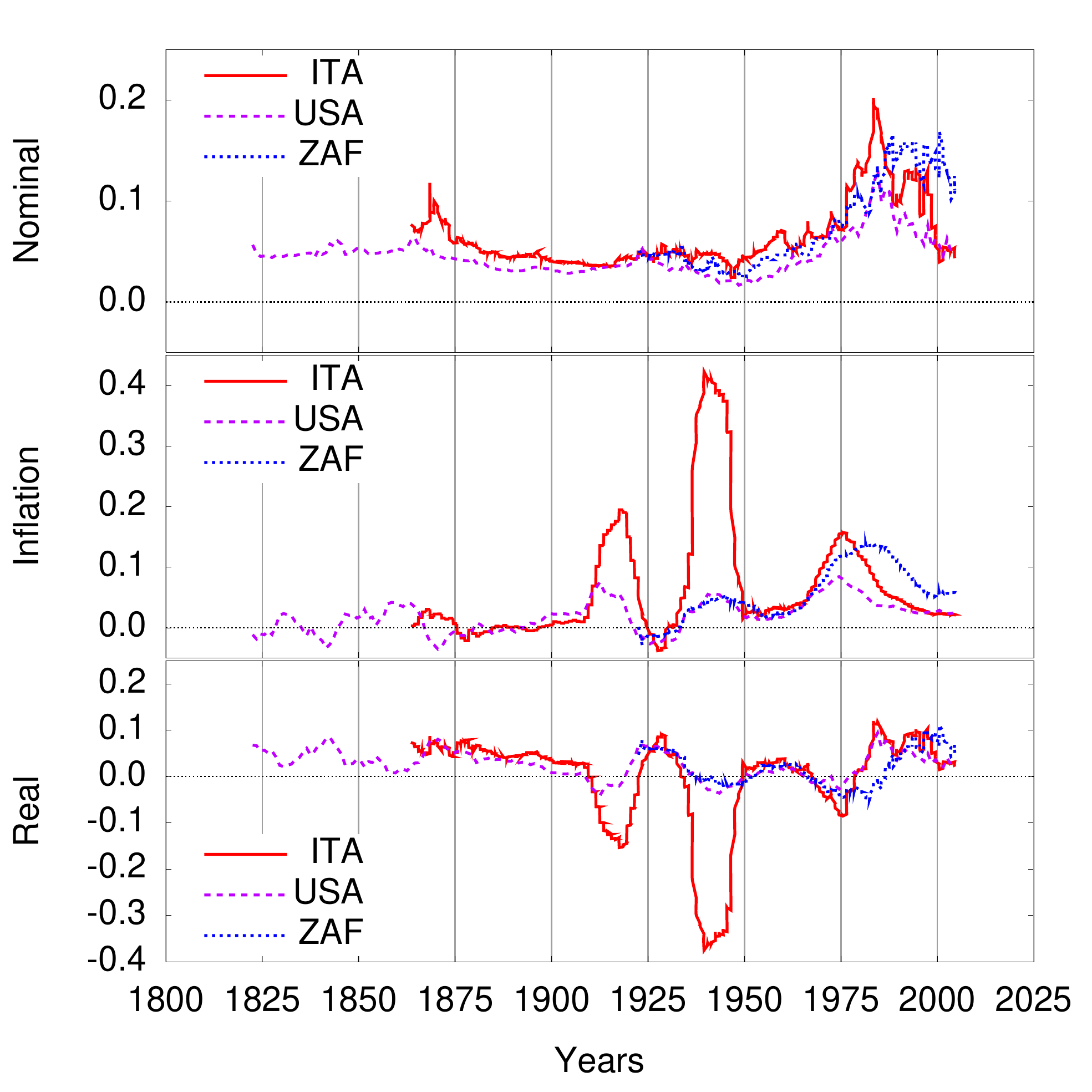}
\caption{{\bf Real interest rates display large fluctuations and negative rates are not uncommon.} We show nominal interest rates (top), inflation (middle), and real interest rates (bottom) for Italy (ITA), United States (USA) and South Africa (ZAF). \label{timeSeries}}
\end{figure}

Finally, the real interest rate $r(t)$ is defined by
\begin{equation}
r(t)=n(t)-i(t).
\label{real_rate_2}
\end{equation}
The recording frequency for each country is either annual or quarterly (see Table \ref{table1}). Some examples of the resulting real interest rates $r(t)$ are plotted in Fig~\ref{timeSeries}.

\section{Choosing the Ornstein-Uhlenbeck model}

A striking feature observed in many epochs for all countries is that real interest rates frequently become negative, often by substantial amounts and for long periods of time (see Fig. \ref{timeSeries} and Table \ref{table2}). This rules out most standard financial models, which assume that interest rates are always positive \cite{Brigo2006}. We thus focus our attention on one of the three most popular stochastic models and on the only one that allows for negative rates: the Ornstein-Uhlenbeck model \cite{Uhlenbeck1930}, also known in the financial and economics literature as the Vasicek model \cite{vasicek} and which is also being used for modelling market volatility \cite{Perello2003,Masoliver2002,Perello2004,Masoliver2006}. The model can be written as \cite{farmer2015}
\begin{equation}
dr(t)=-\alpha(r(t)-m)dt+kdw(t),
\label{dr}
\end{equation}
where $r(t)$ is the real interest rate and $w(t)$ is a Wiener process, a Gaussian process with zero mean and unit variance. The parameter $m$ is a mean value to which the process reverts and coincides with the long-term average of the process (\ref{dr}) :
\begin{equation}
\mathbb E [r(t)]\simeq m.
\label{m}
\end{equation}
The parameter $k$ is expressing the amplitude of the fluctuations and it is related to the variance which in the long-term limit reads
\begin{equation}
\mbox{Var}\left[r(t)\right] \simeq \frac{k^2}{2\alpha}.
\label{var}
\end{equation}
The parameter $\alpha$ is the strength of the reversion to the mean $m$. The autocorrelation function in its long-term limit is
\begin{equation}
K(t-t')=\mathbb E \left[(r(t)-m)(r(t')-m)\right]\simeq \frac{k^2}{2\alpha} e^{-\alpha|t-t'|},
\label{auto}
\end{equation}
where $\alpha^{-1}$ is the correlation time $\tau_c$ as can be seen from the definition  
$$
\tau_c\equiv\frac{1}{K(0)}\int_0^\infty K(\tau)d\tau=\frac 1\alpha.
$$

\begin{table}
\small
\centering
\caption{{\bf Negative rates frequency.} ``Negative RI" and ``Years" give respectively the time ratio and the number of years in which real interest rates are negative.
The last row shows the average over all countries.\label{table2}}
\begin{tabular}{l r r}
\hline\hline
Country & Negative RI & Years  \\ \hline
Argentina	&0.20	&17	\\ 
Australia	&0.23	&33	\\ 
Canada		&0.22	&20	\\ 
Chile		&0.56	&43	\\ 
Denmark		&0.18	&33	\\
Germany 		&0.14	&25	\\
Italy		&0.28	&40	\\  
Japan		&0.33	&26	\\ 
Netherlands	&0.17	&33	\\
South Africa	&0.43	&36	\\ 
Spain		&0.25	&45	\\  
Sweden		&0.28	&38	\\  
United Kingdom	&0.14	&45	\\ 
United States &0.19	&37	\\
 \hline
All countries	&0.26	&34	\\
\hline\hline
\end{tabular}
\end{table}

\begin{figure}
\vspace*{.05in}
\centering
\includegraphics[width=12cm]{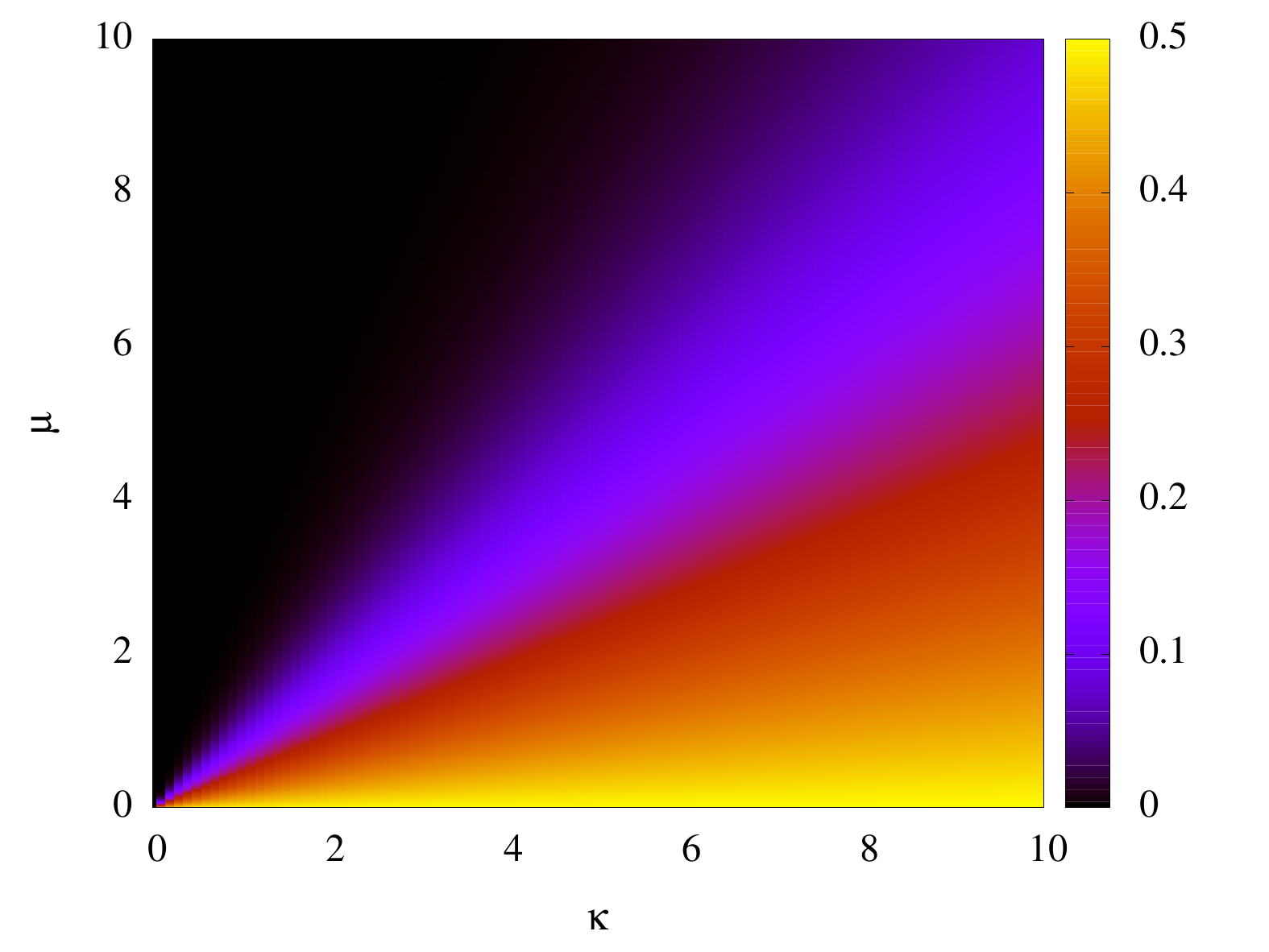}
\caption{The probability of negative rates as given in Eq. (\ref{p-stat2}). In the vicinity of the bottom right corner the probability of negative rates is around $0.5$ while at the upper left corner this probability is exponentially small and rates are mostly positive.\label{pminus}}
\end{figure}

Recall that the OU model may attain negative rates. Let us quantify this characteristic by evaluating the probability $P(r<0,t|r_0)$, for $r(t)$ to be negative. In the long-term limit we denote this probability by $P_s^{(-)}$, that is, 
$$
P_s^{(-)}=\lim_{t\rightarrow\infty}P(r<0,t|r_0).
$$
For the OU model we have
\begin{equation}
P_s^{(-)}=\frac 12{\rm Erfc}\left(\mu/\kappa\right),
\label{p-stat2}
\end{equation}
where ${\rm Erfc}(x)$ is the complementary error function expressed in terms of
\begin{equation}
\mu=\frac{m}{\alpha}, \qquad \kappa=\frac{k}{\alpha^{3/2}}.
\label{mu}
\end{equation}
The dimensionless parameters $\mu$ and $\kappa$ are related to the average $m$ and the noise intensity $k$, respectively. As we will see later, these parameters provide a rather convenient way of describing important features about the discount function $D(t)$. In Fig. \ref{pminus}, we represent Eq. (\ref{p-stat2}) and show the different values that the function $P_s^{(-)}$ can attain in terms of $\mu$ and 
$\kappa$.

Using standard asymptotic expressions of ${\rm Erfc}(x)$ we can also get the behavior of $P_s^{(-)}$ in the cases (i) $\mu<\kappa$ and (ii) $\mu>\kappa$.

(i) When the normal rate $\mu$ is smaller than the volatility of the rate $\kappa$, we can use the series expansion 
$$
{\rm Erfc}(z)=1-\frac{2}{\sqrt\pi}z+O(z^2).
$$
Hence,
\begin{equation}
P_s^{(-)}=\frac 12-\frac{1}{\sqrt\pi}(\mu/\kappa)+O(\mu^2/\kappa^2).
\label{p-i}
\end{equation}
For $\mu/\kappa$ sufficiently small, this probability approaches $1/2$. In other words, rates are positive or negative with almost equal probability. Note that this corresponds to a rather stressed situation in which noise $\kappa$ dominates over the mean value $\mu$.
\noindent

(ii) When fluctuations around the normal level are smaller than the normal level itself, $\kappa<\mu$, we can use the asymptotic approximation
$$
{\rm Erfc}(z)\sim\frac{e^{-z^2}}{\sqrt\pi z}\left[1+O\left(\frac{1}{z^2}\right)\right],
$$
and
\begin{equation}
P_s^{(-)}\sim\frac{1}{2\sqrt\pi}\left(\frac{\kappa}{\mu}\right)e^{-\mu^2/\kappa^2}.
\label{p-ii}
\end{equation}
Therefore, for mild fluctuations around the mean the probability of negative rates is {\it exponentially small}. 

When $\kappa=\mu$, the probability of negative rates is $P_s^{(-)}=0.079$.  Due to the ergodic character of the OU process \cite{Masoliver2018}, this means that when noise is balanced by the mean value (that is, $\kappa=\mu$), one may expect to have negative real rates  $7.9\ \%$ of the time \cite{farmer2015}.

\section{Discount function and negative rates for the Ornstein-Uhlenbeck model}

It is possible to derive the exact expression for the discount function $D(t)$ defined in Eq. (\ref{D}) in the case of the time-dependent OU model. As thoroughly described in Ref. \cite{farmer2015}, we write this expression in the form

\begin{eqnarray}
\ln D(t)=&-&\left(m-\frac{k^2}{2\alpha^2}\right)t \nonumber\\
&+&\frac 1\alpha\left[m-r_0-\frac{k^2}{4\alpha^2}\left(3-e^{-\alpha t}\right)\right]\left(1-e^{-\alpha t}\right).
\label{D1}
\end{eqnarray}

The best way to study the discount rate is to work with the dimensionless time unit $\tau=\alpha t$, for afterwards focussing on the long-term limit $\tau\gg 1$ since climate action is primarily interested in this asymptotic value. Thus, as $\tau \to\infty$, the exact expression (\ref{D1}) shows at once that the discount function of the OU model decays exponentially\footnote{Note also that as $\tau \to 0$ the short-time expansion of Eq. (\ref{D1}) leads to $D(t)\simeq e^{-r_0t}$ which would correspond to a fixed interest rates without random fluctuations or deterministic changes.}
\begin{equation}
D(t)\simeq e^{-r_\infty t},
\label{assymptotic_D}
\end{equation}
where (cf. Eq. (\ref{mu}))
\begin{equation}
r_\infty = m-k^{2}/2\alpha^{2}=\alpha\left(\mu-\kappa^2/2\right).
\label{r_inf}
\end{equation} 

We see from this expression that the long-run discount rate $r_\infty$ is always lower than the average interest rate $m$, by an amount that depends on the  dimensionless noise parameter $\kappa$. The long-run discount rate can therefore be much lower than the mean, and indeed can correspond to low interest rates that are rarely observed. This clearly illustrates the imprudence of assuming that the average real interest rate is the correct long-run discount rate. 

\begin{figure}[tb]
\vspace*{.05in}
\centering
\includegraphics[width=12cm]{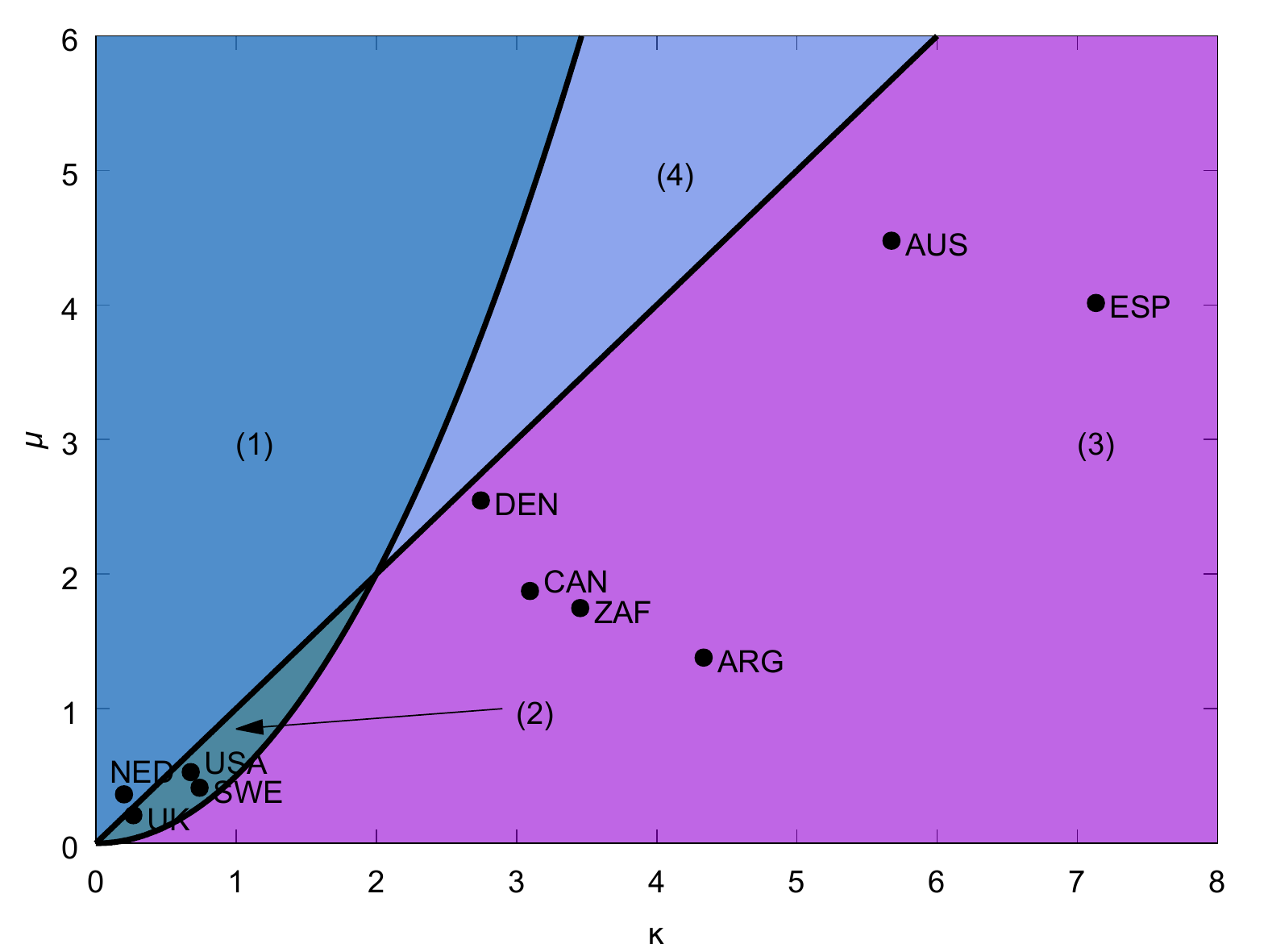}
\caption{{\bf The four different scenarios for the discount with the cases of nine countries.} The vertical axis is the dimensionless mean interest rate $\mu$ and the horizontal axis is the dimensionless fluctuation amplitude $\kappa$. Points correspond to nine of the fourteen countries presented and does not include the errors associated (cf. Table \ref{table4}). The errors are important as can be seen in Table \ref{table4}. Five countries are not reported here because they are far out of the range herein provided.\label{fig:phasePlane}}
\end{figure}

The long-run behavior of the discount rate (\ref{r_inf}) depends on the two dimensionless parameters $\mu$ and $\kappa$ (cf. Eq. (\ref{mu})). The parameter space can be therefore divided into four regions, as shown in Fig.~\ref{fig:phasePlane}. In the region (1), where $\mu>\kappa^{2}/2$ (or equivalently $m>k^{2}/2\alpha^{2}$) and $\mu>\kappa$, the mean interest rate is large in comparison to the noise and negative rates are very infrequent. The long-run discounting function decays exponentially with rate $r_\infty > 0$. In the region (2), albeit small, the long-run discounting function still decays exponentially with rate $r_\infty > 0$ but negative rates are more frequent than $7.9\%$. Region (3) represents the most catastrophic situation since $\mu<\kappa^{2}/2$ and thus $r_\infty < 0$, meaning that the discount function $D(t)$ \textit{increases} exponentially and negative rates are rather frequent. Region (4) also shows $r_\infty < 0$ although, in this case, it is mostly because the noise component is very intense and not due to the presence of a relevant frequency of negative return events. Finally, at the boundary $\mu=\kappa^{2}/2$, the long-run interest rate $r_\infty =0$ and the discount function is asymptotically constant.

\section{Estimating the discount function for the Ornstein-Uhlenbeck model}

\begin{table}
\small
\centering
\caption{{\bf Maximum Likelihood Estimation for the Ornstein-Uhlenbeck process described and the long-run interest rate.} Countries have been reordered based on their estimated $\hat{r}_\infty$. $\hat{m}$ is the estimator of the mean real interest rate in 1/years. $\hat{\alpha}$ is the estimator related to the characteristic reversion time in 1/year. The squared root of the estimator of $k^2$ is the volatility of the process and $k^2$ is given in terms of $1/\mbox{(year)}^3$. These estimators are accompanied with the square root of the variance, $\sigma$'s, of each estimator. $\hat{r}_\infty$ is the subsequent estimator of the long-run real interest rate in 1/year. Negative values of $\hat{r}_\infty$ mean the discount function is asymptotically increasing and its standard error is obtained through error propagation. The last two rows show separately the average over all countries, the stable countries with $r_\infty > 0$ and the unstable countries with $r_\infty < 0$. In all three rows standard error provided corresponds to the standard deviation of the $\hat{r}_\infty$ for the different countries.\label{table3}}
\begin{tabular}{l r r r r r r r r r r r r r r r}
\hline\hline
Country & $\hat{m}$ & $\sigma_{\hat{m}}$ &&$\hat{\alpha}$ & $\sigma_{\hat{\alpha}}$ && $\hat{k^2}$ & $\sigma_{\hat{k^2}}$ && $\hat{r}_\infty$ & $\sigma_{\hat{r}_\infty}$  \\ \hline
Germany 	&-0.0945 &0.6695 &&0.0071	&0.0089	&&41.72E-04 &2.19E-04 &&-40.94 &2.28\\ 
Chile &-0.0579 &0.3146 &&0.0201 &0.0227 &&31.07E-04 &2.49E-04 &&-3.917 &0.442\\ 
Japan &0.0502 &0.2468 &&0.0053 &0.0114	&&13.96E-05 &1.09E-05 &&-2.431 &0.314\\ 
Italy &0.0197 &0.1595 &&0.0056 &0.0089 &&11.46E-05 &0.68E-05 &&-1.778	&0.192\\ 
Spain &0.0671 &0.0692 &&0.0167 &0.0137 &&23.71E-05 &1.26E-05 &&-0.3578	&0.0728\\ 
Argentina	&0.0315 &0.0709 &&0.0228 &0.0231 &&22.40E-05 &1.71E-05 &&-0.1831 &0.0727\\ 
Australia	&0.0397	 &0.0450 &&0.0089 &0.0112 &&2.23E-05 &0.13E-05 &&
-0.1029 &0.0458\\ 
South Africa	&0.0269 &0.0472 &&0.0154 &0.0193 &&4.35E-05 &0.34E-05 &&-0.0649	&0.0477\\ 
Canada		&0.0266 &0.0391 &&0.0142 &0.0178 &&2.75E-05 &0.21E-05 &&-0.0415	&0.0394\\ 
Denmark		&0.0410 &0.0259 &&0.0161 &0.0133 &&3.15E-05 &0.17E-05 &&
-0.0197 	&0.0261\\ 
Sweden		&0.0279	 &0.0166 &&0.0676 &0.0317 &&16.92E-05 &2.06E-05 &&0.0095	&0.0167\\ 
USA			&0.0319 &0.0123 &&0.0603 &0.0257 &&10.03E-05 &1.05E-05 &&
0.0181	&0.0124\\ 
UK			&0.0342	 &0.0062 &&0.1635 &0.0326 &&31.37E-05 &2.53E-05 &&0.0283	&0.0062\\ 
Netherlands	&0.0599 &0.0078 &&0.1648 &0.0550 &&17.97E-05 &2.43E-05 &&0.0566 	&0.0078\\ \hline
All countries &0.0217 &0.1236 &&0.0420	 &0.0211 &&63.45E-05 &4.31E-05 &&-3.552	&0.255\\
Stable		&0.0385 &0.0107 &&0.1140 &0.0362 &&19.07E-05 &2.02E-05 &&0.0281	&0.0108\\ 
Unstable		& 0.0150  &0.1686 &&0.0132 &0.0150 &&81.20E-05 &5.23E-05 &&
-4.984	 &0.353\\
\hline\hline
\end{tabular}
\end{table}


\begin{table}[!ht]
\small
\centering
\caption{{\bf Dimensionless mean interest rate and fluctuation amplitude for all countries.} The dimensionless mean interest rate estimator $\hat{\mu}$ is accompanied with its error obtained through error propagation (cf. Eq. (\ref{mu})) and by considering the parameters estimated and provided in Table \ref{table3}. The dimensionless fluctuation amplitude estimator $\hat{\kappa}$ is accompanied with its standard error obtained through error propagation (cf. Eq. (\ref{mu})) and by considering the parameters estimated and provided in Table \ref{table3}.\label{table4}}
\begin{tabular}{l r r r rr r}
\hline\hline
Country & $\hat{\mu}$ & $\sigma_{\hat{\mu}}$ &&&$\hat{\kappa}$ & $\sigma_{\hat{\kappa}}$ \\ \hline
Germany 	& -13.22 &95.11 &&&106.92	&198.75\\
Chile 	&-2.89	&16.01	&&&19.61	&33.26\\
Japan	&9.46	&50.79	&&&30.59	&98.70\\
Italy	&3.49 &28.79 &&&25.23 &59.94\\
Spain	&4.02	&5.30	&&&7.13	&8.79\\
Argentina&1.38	&3.40	&&&4.34	&6.58\\
Australia&4.48	&7.61 &&&5.67 &10.77\\
South Africa	&1.75 &3.77	&&&3.45 &6.50\\
Canada&1.88	&3.62	&&&3.10 &5.83\\
Denmark&2.55 &2.65 &&&2.75	&3.41\\
Sweden&0.41	&0.31	&&&0.74	&0.52\\
USA	&0.53	&0.30	&&&0.68	&0.43\\
UK	&0.21	&0.06	&&&0.27 &0.08\\
Netherlands&0.36	&0.13 &&&0.20 &0.10\\ \hline
All countries&1.03	&15.56 &&&15.05	&30.99\\
Stable&0.39	&0.20 &&&0.47	&0.28\\
Unstable&1.29	&21.71	&&&20.89 &43.25\\ 
\hline\hline
\end{tabular}
\end{table}

We now estimate the parameters $m$, $k$ and $\alpha$ together with the dimensionless parameters $\mu$ and $\kappa$ defined in Eq. (\ref{mu}). We perform such an estimation for each historical series (cf. Table \ref{table1}) by using a well-established maximum likelihood procedure for the OU model \cite{Brigo2006}. The resulting estimators $\hat{m}$, $\hat{\alpha}$, and $\hat{k^2}$ are listed in Table \ref{table3} along with their standard deviation derived from formulas provided in Ref. \cite{Chen2009}. Table \ref{table3} shows that the most inaccurate estimator is $\hat{\alpha}$, a not surprising fact since the estimation of $\alpha$ is quite a challenge in any Ornstein-Uhlenbeck process \cite{Chen2009}. The last two columns in Table \ref{table3} include the long-run interest rate estimator $\hat{r}_\infty$ and its error calculated through error propagation.

We can also observe the position $(\hat{\kappa}, \hat{\mu})$ of each country in Fig.~\ref{fig:phasePlane} by considering the results presented in Table \ref{table4}. In any case these results need to be understood as a first-order approximation since the errors behind the estimators (which are evaluated  through error propagation) are significant (see Table \ref{table4}). 
Only four countries show a positive long-run rate, $r_{\infty} > 0$, and all of them inside, or very close, to the region defined by 
$\mu<\kappa$ in which rates are frequently negative. The other ten countries show less stable behavior and are all of them in the exponentially increasing region (Region 3), which implies they have long-run negative rates, and are widely scattered. In two cases (Germany and Chile) the average rate $m$ (and its dimensionless version $\mu$) is negative due to at least one period of runaway inflation while two others (Japan and Italy) still have a long-run negative rate $r_{\infty}$ mostly due to a very small strength of the reversion to the mean given by the parameter $\alpha$ (cf. Eq. (\ref{dr})). These four countries are not plotted in Fig. \ref{fig:phasePlane} because they are out the range of $\mu$ and/or $\kappa$ axis. 

Also note that all fourteen countries but one (Netherlands) are below the identity line, $\mu=\kappa$,  in Fig~\ref{fig:phasePlane} which indicates that negative real interest rates are common (even in the stable countries they occur $20\%$ of the time). It is also worth to mention that only one is above Nordhaus's 4$\%$ discounting rate \cite{Nordhaus} (5.7$\%$, Netherlands) and only two more countries are above the more pessimistic discounting rate (1.4$\%$) provided by Stern \cite{Stern} (1.8$\%$ and 2.8$\%$ from USA and United Kingdom, respectively). And more generally, it is important to notice that $r_{\infty}$ is very much smaller than $m$ in most of the cases. All these statements are robust even when considering values of the estimators with shifts of the size of its standard error (see Table \ref{table3}). 

\begin{figure}
\vspace*{.05in}
\centering
\includegraphics[width=14cm]{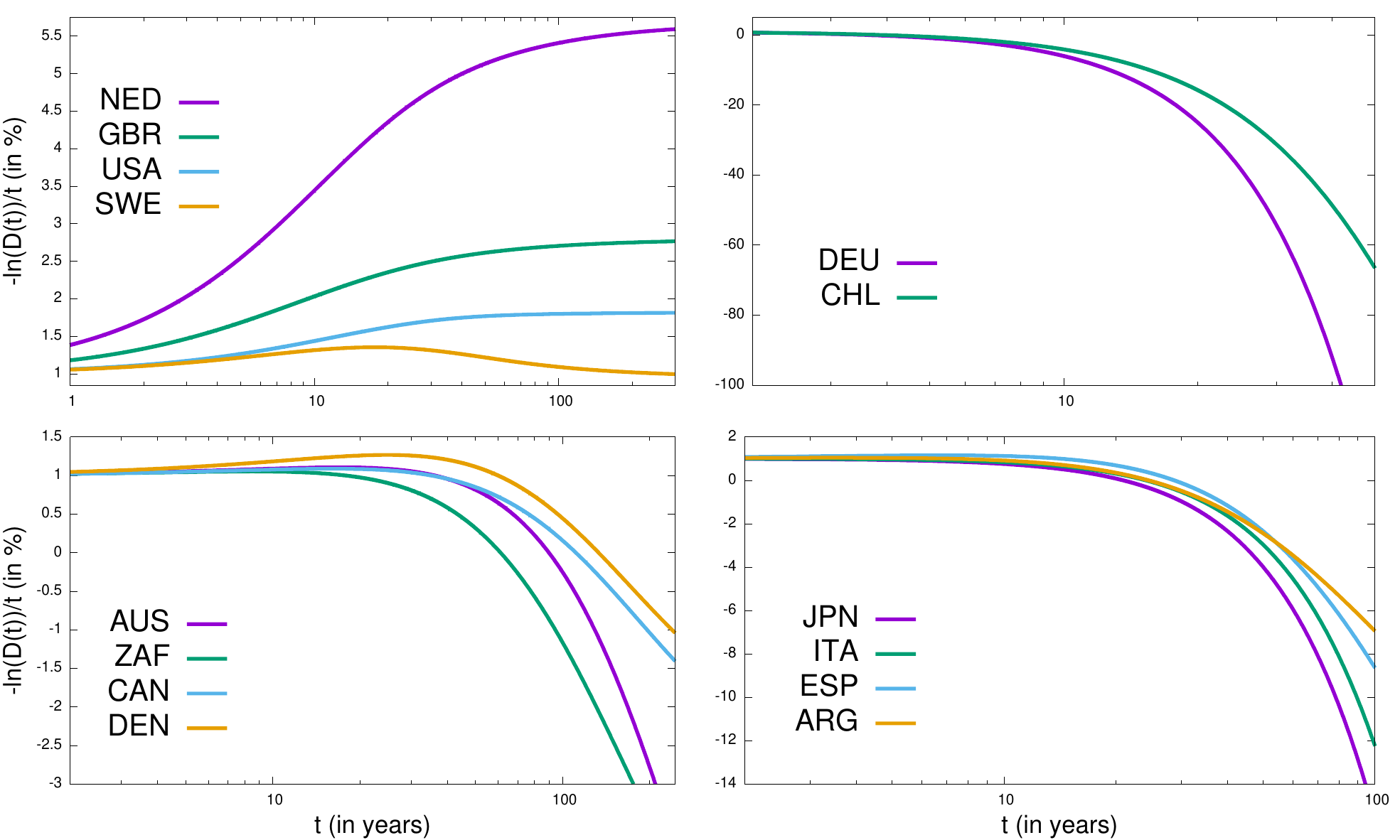}
\caption{{\bf The logarithmic discounting rate (in percent) as a function of time (in years).} We have divided the countries in four groups to represent Eq. (\ref{D1}) with parameters provided in Table \ref{table4} and taking $r_0=1\%$.\label{discountExamples}}
\end{figure} 

\begin{figure}
\vspace*{.05in}
\centering
\includegraphics[width=14cm]{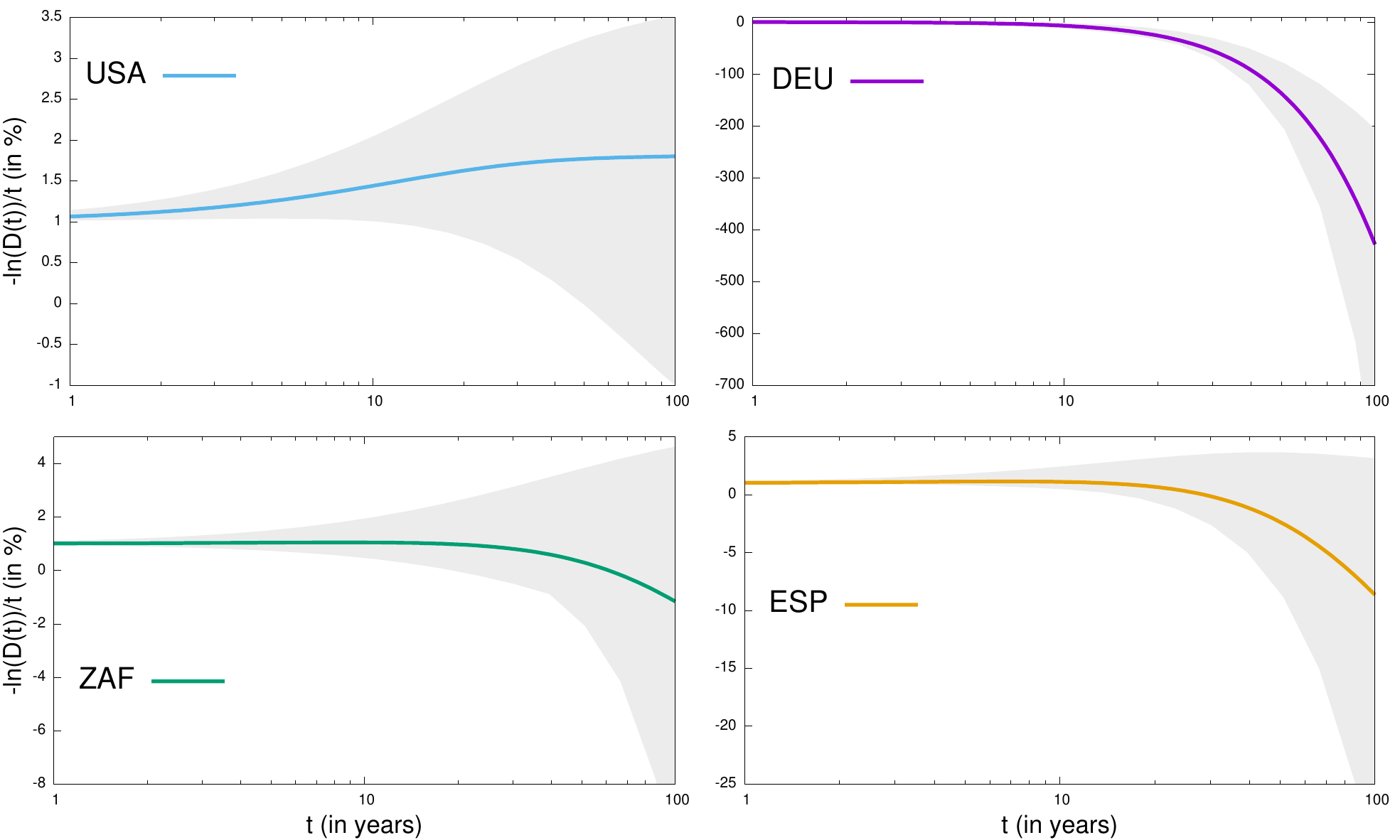}
\caption{{\bf The logarithmic discounting rate (in percent) and its error as a function of time (in years).} We have selected four countries (from each of the four groups provided in Fig. \ref{discountExamples}: United States, Germany, South Africa, and Spain) to represent Eq. (\ref{D1}) by taking $r_0=1\%$. Grey shadow considers a range limited by minimum and maximum values when adding to each parameter the impact of their standard error to the value of the discount rate as a function of time. Parameters and their standard error are both provided in Table \ref{table3}.\label{discountExampleserror}}
\end{figure} 

The characteristic (correlation) time ($\tau_c=1/\alpha$) for each country appears to be very different (cf. Table \ref{table3}). Some countries must spend more than a century to achieve a stationary level and thus finally attain the long-run discount rate $r_{\infty}$. Furthermore, this time horizon might be even larger than the time interval we must consider to make a response, from an economic point of view, to any climate change catastrophe. For this reason, it is interesting to investigate how the discount rate defined as 
$-\ln(D(t))/t$ changes over time (cf. Eq. (\ref{D1})). 

Figure~\ref{discountExamples} shows the discount rates for all countries as a function of time by considering initial rate $r_0=1\%$ which clearly illustrates the dramatic differences between countries. In this way we divide the fourteen countries into roughly four groups. There are two countries (DEU, CHL) that show a very fast and very negative rate. There is a second group still having a monotonic behavior but with a much slower trend to raise negative discount rates (JPN, ITA, ESP and ARG). Non-monotonic behavior is indeed observed in a third group (AUS, ZAF, CAN, DEN). This group is of special interest since it shows how the rates might first grow by finally becoming negative after 20 or 30 years. Stable countries represented in the first inset on the left of Fig. \ref{discountExamples} also show that the asymptotic rate $r_{\infty}$ is raised very slowly being the country with the highest rate (NED) the one that needs more than a century to attain the stationary level. Figure \ref{discountExampleserror} selects four countries (USA, DEU, ZAF and ESP), one from each of the groups mentioned above, to observe the impact of uncertainty as a function of time. For different values of time, the discount function includes a shadow in grey limited by maximum and minimum values when taking into consideration the standard error of each of the estimators. In all four countries and at any time, maximum discount rate value is always below 2.2\%. The inclusion of the statistical uncertainty reinforces the robustness of our results.

Let us finally note that these results are in contrast to other treatments of fluctuating rates which assume that short term rates are  positive and predict that the decrease in the discounting rate occurs over a much longer timescale, usually  measured in hundreds or thousands of years \cite{Weitzman98,NewellPizer,Groom,Farmer,Hepburn,Freeman,GollierBook}.

\section{Considering alternative models}

As mentioned above the Ornstein-Uhlenbeck model is the only one among the three most classic models allowing for negative rates. This is the reason why we have excluded both the Feller and the lognormal models from our analysis. Let us nonetheless briefly study what modifications should be carried out in order to use these positive rate models in our analysis. 

\subsection{The shifted Feller model}

The Feller process \cite{Feller1951} (see also \cite{Masoliver2018}) has a very similar structure than the Ornstein-Uhlenbeck process except that the noise component depends on the interest rate. The process also has a mean reverting force that makes the process have an autocorrelation function that decays in an exponential manner whose characteristic time scale is $1/\alpha_F$. Let us however remind that Feller does not allow for negative rates and these are clearly present in our empirical data. Therefore, to consider the Feller process for estimating the long-run discount rate $r_{\infty}$ would require to redefine the Feller model that reads
\begin{equation}
dy(t)=-\alpha_F(y(t)-m_F)dt+k_F\sqrt{y(t)}dw(t),
\label{fellermod}
\end{equation}
where 
\begin{equation}
y=r-r_{min}
\label{y}
\end{equation}
and $r_{min}<0$ is the minimum value observed in the time series. The estimation through maximum likelihood procedure and its error analysis is also possible \cite{Chen2009}. Table \ref{table5} compares the Ornstein-Uhlenbeck and the shifted Feller models which have very similar mathematical expressions for estimating $\hat{m}$, $\hat{\alpha}$ and $\hat{k}^2$ parameters. The discount function now reads (cf. Eqs. (\ref{D})) and (\ref{y}))
$$
D(t) = \mathbb E \left[ \exp{\left(-\int_{0}^{t}\left(r_{min}+y(t^{\prime})dt^{\prime}\right)\right)}\right]=\exp\left(-r_{min}t\right)\mathbb E \left[\exp{\left(-\int_{0}^{t} y(t^{\prime})dt^{\prime}\right)}\right].
$$
The asymptotic value of the remaining average shows an exponential decay \cite{farmer2015}
$$
\mathbb E \left[\exp{\left(-\int_{0}^{t} y(t^{\prime})dt^{\prime}\right)}\right]\simeq e^{-y^F_{\infty} t},
$$
whose long-run discount rate \cite{farmer2015} is
$$
y^F_\infty=\frac{2m_F}{1+\sqrt{1+2k_F^2/\alpha_F^2}},
$$
so that 
$$
D(t) =e^{-\left(r_{min}+y^F_{\infty}\right)t}=e^{-r^F_{\infty}t}.
$$
We observe that (as in the Ornstein-Uhlenbeck process) the log-run rate is smaller than the average rate, $r^F_\infty<m$. However, the shifted Feller process leads us to obtain a slightly larger estimation but within the statistical error range ($3.4\%$ versus $1.81\%$, see Table \ref{table5}). The value is similar  than  Nordhaus's $4\%$ discounting rate if one considers the statistical error and in any case lower than $6\%$ \cite{Nordhaus,Nordhaus2007}.

\begin{table}
\small
\centering
\caption{{\bf Maximum Likelihood Estimation for the three different models described and the long-run interest rate for the case of United States.} These estimators are taking years as a basic time unit and they are all accompanied with the square root of the variance, $\sigma$'s, of each estimator. $\hat{r}_\infty$ is the subsequent estimator of the long-run real interest rate in 
1/year. For the Feller and lognormal cases we have provided a modified version of the model (cf. Eqs. (\ref{fellermod}) and (\ref{lognormalmod})). The shifted Feller and lognormal models takes $r_{min}=-0.0415$, which is its minimum value in the historical time series, and the estimated $\hat{r}_{\infty}$ is corrected by adding $r_{min}$ but this has been impossible to be done in the lognormal case since the asymptotic value goes to a constant.\label{table5}}
\begin{tabular}{l r r r r r r r r r r r r r r r}
\hline\hline
 & $\hat{m}$ & $\sigma_{\hat{m}}$ &&$\hat{\alpha}$ & $\sigma_{\hat{\alpha}}$ && $\hat{k}^2$ & $\sigma_{\hat{k}^2}$ && $\hat{r}_\infty$ & $\sigma_{\hat{r}_\infty}$  \\ \hline 
Ornstein-Uhlenbeck	&0.0319 &0.0123 &&0.0603 &0.0257 &&10.03E-05 &1.05E-05 &&
0.0181	&0.0124\\
\\
 & $\hat{m}_F$ & $\sigma_{\hat{m}_F}$ &&$\hat{\alpha}_F$ & $\sigma_{\hat{\alpha}_F}$ && $\hat{k}_F^2$ & $\sigma_{\hat{k^2}_F}$ && $\hat{r}^F_\infty$ & $\sigma_{\hat{r}^F_\infty}$  \\ \hline
Shifted Feller & 0.0864 & 0.0041 && 0.0599 & 0.0057 && 12.56E-05 &  1.31E-05 && 0.0349 & 0.0072 \\ \\
 & $\hat{m}_L$ & $\sigma_{\hat{m}_L}$ && $\hat{k}_L^2$ & $\sigma_{\hat{k}_L^2}$ && $\hat{m}_L-\hat{k}_L^2/2$ & $\sigma_{\hat{m}_L-\hat{k}_L^2/2}$ && Asymp  \\
\hline
Shifted lognormal & 0.0130 & 0.0163 && 0.0309 & 0.0066 && -0.0024 & 0.0130 && \mbox{constant}  \\
\hline\hline
\end{tabular}
\end{table}

\subsection{The shifted lognormal model}

Another alternative to still allow for negative rates is to consider a modified version of the lognormal process by considering new variable $y=r-r_{min}$ (where $r_{min}<0$ is the minimum value observed in the time series) and the following stochastic dynamics:
\begin{equation}
dy(t)=m_Ly(t)dt+k_Ly(t)dw(t),
\label{lognormalmod}
\end{equation}
whose long-run discount function can lead to three different asymptotic expressions \cite{farmer2015}:
$$
\mathbb E \left[\exp{\left(-\int_{0}^{t} y(t^{\prime})dt^{\prime}\right)}\right]\sim 
\begin{cases}
\mathrm{constant} & \quad m_L<k_L^2/2,\\
e^{-y^L_{\infty}  t} & \quad m_L>k_L^2/2.\\
t^{-1/2} & \quad m_L=k_L^2/2,
\end{cases}
$$
For the exponential case the long-run discount rate reads
$$
y^L_{\infty} =\frac{m_L-k^2_L/2}{\psi\left(2m_L/k_L^2\right)+1/(2m_L/k_L^2-1)},
$$
where $\psi(\cdot)$ is the digamma function. The lognormal process does not show any reversion trend to a certain level and its average grows (or decreases) in an exponential manner
$$
\mathbb{E}\left[r(t)|r(0)=r_0\right]=(r_0+r_{min}) e^{m_L t}-r_{min},
$$
a result that it is in contradiction with the times series provided in Fig. \ref{timeSeries}. We can however also estimate the parameters via maximum likelihood procedures. The results are again provided in Table \ref{table5} and they show us that the asymptotic discount falls into the constant case since $m_L<k_L^2/2$ although the error analysis warn us that we cannot discard the exponential case (being not greater $4-5\%$) nor the hyperbolic slow decay.

\subsection{Extending the Ornstein-Uhlenbeck process} 

One can argue that the results presented can change under different historical conditions or periods. To exemplify this issue, we have also estimated these values in the case of Germany once the World War II was over (from March 1946). Parameters are in that case $\mu=0.62$, $\kappa=0.32$ with now a positive long-run rate $r_\infty=3.4\%$ which is in any case smaller than Nordhaus estimates for valuing climate action \cite{Nordhaus2007}. Germany certainly is a quite volatile situation  challenging the model which assumes constant (i.e., stationary) parameters. A possible way out is to extend the model with an additional dimension under the form of a ``matrioshka doll'' by considering $m$ as a stochastic process following an additional Ornstein-Uhlenbeck process \cite{Perello2004}\footnote{The model thus consists of two Ornstein-Uhlenbeck processes one inside the other. Hence the name ``matrioshka doll''.}
\begin{eqnarray}
&&dr=-\alpha(r-m)dt+kdw(t) \nonumber \\
&&dm=-\alpha_0(m-m_0)dt+k_0dw_0(t),
\label{doubleou}
\end{eqnarray}
where the Wiener processes $w(t)$ and $w_0(t)$ are both zero mean, have unit variance and are independent from each other implying that $\mathbb E \left[dw(t) dw_0(t)\right]=0$. We also assume that $\alpha>\alpha_0>0$ thus showing a slower mean reverting force for the subordinated process $m_0$ than for $m$. A similar extension has been used in other financial contexts to model stochastic volatility \cite{Perello2004} by adding a longer mean reversion process which allows for a slow decaying memory for the volatility process while still preserving a much shorter memory for the so-called leverage effect \cite{Masoliver2006,Perello2003} (see also Ref. \cite{Delpini2015} for another setting that could represent an alternative approach to the extended Ornstein-Uhlenbeck model given by Eq. (\ref{doubleou})). In the long-run, the process reads \cite{Perello2004}:
\begin{eqnarray}
m(t)&=&m_0+k_0\int_{-\infty}^t e^{-\alpha_0(t-t')}dw_0(t) \nonumber \\
r(t)&=&m_0+k\int_{-\infty}^t e^{-\alpha(t-t')}dw(t)+\frac{k_0}{\alpha-\alpha_0}\int_{-\infty}^t \left(e^{-\alpha_0(t-t')}-e^{-\alpha(t-t')}\right)dw_0(t).
\label{doubleouint}
\end{eqnarray}
We can easily see that this extended process shows the same average, $\mathbb E \left[r(t)\right]=m_0$, than the simpler OU version but with greater variance (cf. Eqs. (\ref{m}) and (\ref{var}), respectively)
\begin{equation}
{\rm Var}[r(t)]=k^2/2\alpha+k_0^2/2\alpha_0
\label{varextended}
\end{equation}
The autocorrelation function now reads \cite{Perello2004} (cf. Eq. (\ref{auto})):
\begin{eqnarray*}
K(t-t')&=&\mathbb E \left[(r(t)-m_0)(r(t')-m_0)\right]\nonumber \\
&=&\left(\frac{k^2}{2\alpha}-\frac{k_0^2\alpha}{2(\alpha^2-\alpha_0^2)}\right)e^{-\alpha|t-t'|}+\frac{k_0^2\alpha^2}{2(\alpha^2-\alpha_0^2)\alpha_0} e^{-\alpha_0|t-t'|}, \label{autodouble}
\end{eqnarray*}
where the first term with an exponential decay with $1/\alpha$ would dominate for short time difference $|t-t'|$. In the opposite situation, for longer time difference, the second exponential decay expressed in terms of $1/\alpha_0$ would dominate. The extended process now have five parameters ($\alpha,k,\alpha_0,k_0$, and $m_0$), while basic OU process had only three ($\alpha,k$, and $m$).

We can finally look at the effects on the asymptotic discount. It can be in this case proved that the process has also an exponential decay with a long-run discount rate that reads \cite{Masoliver2020}
\begin{equation}
r_{\infty}^{ext}=m_0-\frac{1}{2}\left(\frac{k^2}{2\alpha^2}+\frac{k_0^2}{2\alpha_0^2}\right).
\end{equation}
The result brings rates which are even lower than the one provided by the maximum likelihood estimation procedure in the simple Ornstein-Uhlenbeck process. As a simple exercise we can estimate a combination of $k_0$ and $\alpha_0$ with the historical variance of the whole process (see Eq. (\ref{varextended})). To estimate $\alpha_0$ is not that simple since our historical data sets are too short and its estimation becomes too noisy. However, jointly with the values already obtained for Orstein-Uhlenbeck maximum likelihood estimation for $m$ (now equivalent to $m_0$), $k$, and $\alpha$, it is possible to observe the effects for different values of $\alpha_0$:
$$
r_{\infty}^{ext}=m_0-\frac{1}{2}\left[\frac{k^2}{2\alpha^2}+\left({\rm Var}(r(t))-\frac{k^2}{2\alpha}\right)\frac{1}{\alpha_0}\right].
$$
In this case we can see that the long-run rate $r_\infty$ for the United States is practically zero when 
$\alpha_0=\alpha/20$. 

\section{Discussion}

Our empirical analysis proves that real interest rates are often negative --roughly a quarter of the time-- which implies that  one must use a discount model that is compatible with this property. For this purpose we have proposed the Ornstein-Uhlenbeck model which has the additional advantage that it can be solved analytically in a relatively simple way. This model facilitates the understanding of why the long-run discount rate can be so low. A first reason is that real interest rates are themselves typically low. As we have  showed the average over all countries surveyed is negative, and even the average over stable countries (those with a positive long-run rate, $r_\infty > 0$) is $2.8\%$. A second reason is that the fluctuating part on the right hand side of Eq.\ (\ref{r_inf}), which depends both on the noise intensity $k$ and the persistence term $1/\alpha$, typically lowers rates for the stable countries by about $7\%$. In some cases, such as Spain, the effect is much more dramatic: Even though the mean short term rate has the high value of $m = 6.7\%$, the long-term discounting rate is $r_\infty = -36\%$ which would imply a great increasing discount. The estimation is being done with a maximum likelihood procedure that includes an error analysis that demonstrates the robustness of the results obtained.

Our analysis here makes several simplifications such as ignoring non-stationarity. We have here partially address this issue by extending the Ornstein-Uhlenbeck which allows for slower fluctuations in the normal level and resulting even in lower  long-run discount rates \cite{Masoliver2020}. Correlations between the environment and the economy have also being ignored but, in any case, despite the variety of results, the long-run discount rate is always smaller than Nordhaus estimates by other methods as we have exemplified with the German case \cite{Nordhaus2007}. We have also not considered the market price of risk \cite{vasicek,masoliver_2018(b)}, in other words, we have assumed that markets are risk neutral and the average in Eq. (\ref{D}) defining the discount function, is evaluated using the empirical probability measure without any risk adjustment \cite{cox_81,piazzesi}. These issues are under present investigation and some results are expected soon \cite{farmer2020}.

In any case the methods that we have introduced here provide a foundation on which to incorporate more realistic assumptions. We do not mean to imply that it is realistic to actually use the increasing discounting functions that occur for countries with less stable interest rate processes. There is some validity to treating hyper-inflation as an aberration -- when it occurs government bonds are widely abandoned in favor of more stable carriers of wealth such as land and gold, and as a result under such circumstances the difference between nominal interest and inflation may underestimate the actual real rate of interest.

Nonetheless, real interest rates are closely related to economic growth, and economic downturns are a reality. The great depression lasted for 15 years, and the fall of Rome triggered a depression in western Europe that lasted almost a thousand years. In light of our results here, arguments that we should wait to act on global warming because future economic growth will easily solve the problem should be viewed with extreme skepticism. Our analysis clearly supports Stern over Nordhaus. When we plan for the future we should always bear in mind that sustained economic downturns may visit us again, as they had in the past. 

Effective responses to this multifaceted problem have been slow to develop, in large part because many experts have not only underestimated its impact, but also overlooked the underlying institutional structure, organizational power and financial roots \cite{Stern2016,Farrell2019}. A growing body of sophisticated research is currently emerging with a large set of multidisciplinary strategies that wants to exploit socioeconomic tipping points (as in any complex dynamical system) to magnify the impact of each political intervention \cite{Farmer2019} and also integrate science-policy perspectives with public awareness, citizen-led research and citizen science practices (see for instance \cite{Vicens2019,Kythreotis2019}). In all cases the final purpose is to better respond to global challenges such as climate action in a near future, sooner rather than later.

\subsection*{Acknowledgements}

We thank to an anonymous referee for the several comments that have allowed to improve the final version of the paper. We also would like to thank National Science Foundation grant 0624351 (JG) and the Institute for New Economic Thinking (JDF). This work was supported by MINECO (Spain) FIS2013-47532-C3-2-P (JM, MM and JP), FIS2016-78904-C3-2-P (JM, MM and JP); by Generalitat de Catalunya (Spain) through Complexity Lab Barcelona (contract no. 2017 SGR 1064, JM, MM and JP)




%
%
%

\end{document}